# Genome Sizes and the Benford Distribution

James L. Friar[1], Terrance Goldman[1], Juan Pérez–Mercader[2]*

1 Theoretical Division, Los Alamos National Laboratory, Los Alamos, New Mexico, United States of America, 2 Department of Earth and Planetary Sciences, Harvard University, Cambridge, Massachusetts, United States of America and Santa Fe Institute, Santa Fe, New Mexico, United States of America

**Abstract**

*Background:* Data on the number of Open Reading Frames (ORFs) coded by genomes from the 3 domains of Life show the presence of some notable general features. These include essential differences between the Prokaryotes and Eukaryotes, with the number of ORFs growing linearly with total genome size for the former, but only logarithmically for the latter.

*Results:* Simply by assuming that the (protein) coding and non-coding fractions of the genome must have different dynamics and that the non-coding fraction must be particularly versatile and therefore be controlled by a variety of (unspecified) probability distribution functions (pdf's), we are able to predict that the number of ORFs for Eukaryotes follows a Benford distribution and must therefore have a specific logarithmic form. Using the data for the 1000+ genomes available to us in early 2010, we find that the Benford distribution provides excellent fits to the data over several orders of magnitude.

*Conclusions:* In its linear regime the Benford distribution produces excellent fits to the Prokaryote data, while the full non-linear form of the distribution similarly provides an excellent fit to the Eukaryote data. Furthermore, in their region of overlap the salient features are statistically congruent. This allows us to interpret the difference between Prokaryotes and Eukaryotes as the manifestation of the increased demand in the biological functions required for the larger Eukaryotes, to estimate some minimal genome sizes, and to predict a maximal Prokaryote genome size on the order of 8–12 megabasepairs.These results naturally allow a mathematical interpretation in terms of maximal entropy and, therefore, most efficient information transmission.

**Funding:** This work was carried out in part under the auspices of the National Nuclear Security Administration of the United States Department of Energy at Los Alamos National Laboratory under Contract No. DE-AC52-06NA25396. JP-M would like to thank the everis Foundation and Repsol S. A. for support, and the Theoretical Division of Los Alamos National Laboratory for its hospitality. The funders had no role in study design, data collection and analysis, decision to publish, or preparation of the manuscript.

**Competing Interests:** The authors have declared that no competing interests exist.

* E-mail: jperezmercader@fas.harvard.edu

## Introduction

A substantial number of genomes from all three domains of Life have been sequenced in the past few years. Quality data on many of the individual properties of these genomes are now available. Amongst the data is the total number of base pairs in each genome (or genome size, $G$), which (for a number of cases) has been further broken down into the number of "coding" base pairs, $cDNA = c$, and "non-coding" base pairs, $ncDNA = nc$. In addition, the number of Open Reading Frames (ORFs) in the genome has also been tabulated. Both quantities are important in assessing the functional complexity of the living system.

The relationship between genome size, $G = cDNA + ncDNA \equiv c + nc$ and the number of ORFs in a genome, $y_{ORF}$, is particularly interesting for many reasons that have to do with genome structure, as well as a potential connection with the regularities in the complexity of the organism. These properties are believed to apply across all three domains of Life on Earth, and have been extensively discussed in the literature, for example in connection with genome evolution and architecture [1] or as a means to explore and unravel the probability distribution functions that control the dynamics of a genome, as in Ref. [2].

In this paper we report on the results of a phenomenological study of the relationship between $y_{ORF}$ and $G$ (viz., $y_{ORF}(G)$) for genomes from the three domains of Life. We base our work on the genomic data that were openly available [3] in early 2010. These data are plotted in Figure 1 for each domain.

The paper is organized into this Introduction followed by Results, Discussion, Methods and Acknowledgments sections.

### 1. The Various Forms of the Benford (or Reciprocal) Distribution with an Eye towards its Application to $y_{ORF}(G)$

We will be dealing in the Results Section with the Benford [4] (or reciprocal [5,6]) probability distribution function (pdf). This distribution is central to this paper, and we introduce it here together with a necessarily brief discussion of several aspects (including some specific to our use in the biological context). There are two aspects of the Benford distribution (whose pdf has the form $p(\sigma) \propto 1/\sigma$, where $\sigma$ is the stochastic variable; it is from this algebraic form that derives its alternative name of the "reciprocal distribution") that present formal problems: (1) it diverges as $\sigma \to 0$, and (2) it does not accommodate the fact that living systems cannot exist for arbitrarily small genome size. (In other words, there exists a minimum genome size below which a chemical system will not "boot-up" as a living system. We will denote this genome size by $G^{(min)}$.) These difficulties can be formally avoided by either introducing a cut–off in the denominator of the standard Benford pdf, as will be done in Eqn. (9), or by choosing appropriate limits of integration over the Benford pdf. Both procedures must yield the same result for $y_{ORF}(G)$, and we demonstrate this next with the standard notation used for "the law of significant digits", which is the oldest and best-known application of Benford's law [4] (see also the last reference in [7–9]).

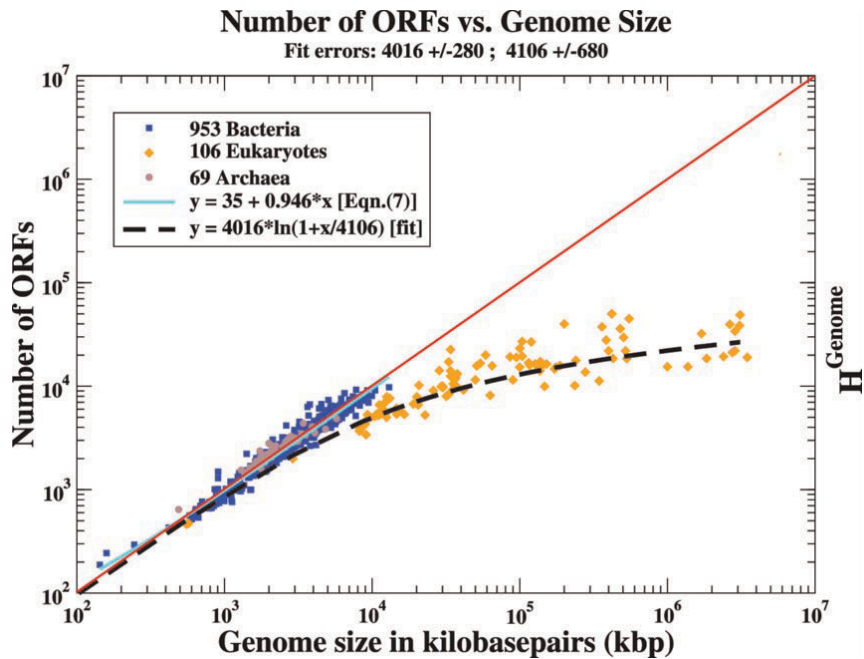

**Figure 1. The Number of ORFs in Each Genome vs. Genome Size for the Three Extant Domains of Life on Earth.** The points are data from 1128 genomes available on the GOLD database [3] in early 2010. In this log-log plot, the x-axis represents the genome size (*G*) in kilobasepairs. For each genome we plot on the y-axis the number of ORFs quoted for the genome in the above database. In order to facilitate comparisons, we have drawn a red diagonal line on a vertical/horizontal scale where 1 vertical axis unit corresponds to 1 kbp on the horizontal axis. The Prokaryotic genomes cluster around this (slope = 1) line. The fit to the Prokaryotes given by Eqn. (6) is represented here as a cyan line. The dashed line represents the best fit to the Eukaryotic ORFs and corresponds to a Benford distribution, Eqn. (11), if we neglect the statistically insignificant contribution from the combination of the first two terms, $y_{ORF}^{(min)} - A \cdot \ln(1 + G^{(min)}/B)$. Note the wide range of genome sizes that the fit accommodates. See the Discussion Section regarding the right-hand axis.
doi:10.1371/journal.pone.0036624.g001

Assuming that the Benford distribution applies to $y_{ORF}$ as a function of $G$, we can formally write $y_{ORF}(G)$ (for $G \geq G^{(min)}$) as an integral between appropriate limits of the Benford pdf:

$$y_{ORF}(G) = y_{ORF}(G^{(min)}) + A \int_{B+G^{(min)}}^{B+G} \frac{d\sigma}{\sigma}, \quad (1)$$

where $B$ is the same parameter introduced later in Eqn. (9) and $y_{ORF}(G^{(min)})$ is the number of ORFs corresponding to the assumed minimal genome size, $G^{(min)}$, and physically measures the minimal number of ORFs needed to "boot-up" a minimal living system (L/S). (Unfortunately, we are not able to provide any details on the **nature** of these ORFs or on how they operate.) Note that the integral vanishes for $G = G^{(min)}$ (as it must for consistency) and there is no problem for small values of $\sigma$ if either $B$ or $G^{(min)}$ is non-vanishing. Equation (1) (with $G^{(min)}$ set to zero) can be recognized as the basic form that is used to derive Benford's Law (the "law of significant digits" [4]) from the eponymous distribution [10].

Performing the integral in Eqn. (1) and using the notation $y_{ORF}(G^{(min)}) \equiv y_{ORF}^{(min)}$ leads immediately to

$$y_{ORF}(G) = \left\{y_{ORF}^{(min)} - A \cdot \ln(1 + G^{(min)}/B)\right\} + A \cdot \ln(1 + G/B), \quad (2)$$

which is Eqn. (11) of the paper. This establishes the equality of the two ways of parametrizing the Benford distribution.

## 2. Some Selected Properties of the Benford (or Reciprocal) Distribution

In this paper we make use of several known properties of the Benford distribution and we list them here for convenience. Their proof and discussion as well as further references can be found in Refs. [5–11].

These properties are the following:

(a) if the sizes of a stochastic variable, $s(k)$, are classified by their rank $k$ in that distribution and are geometrically distributed, then the pdf of that stochastic variable as a function of the size, $p(s)$, must be the Benford distribution;

(b) the errors generated during the combination of quantities distributed according to the reciprocal distribution are smaller than the ones generated if the quantities were uniformly distributed;

(c) the random combination of stochastic quantities selected from a stochastic combination of pdf's **produces** a stochastic variable that is distributed according to the Benford distribution, and

(d) if one combines two stochastic variables through independent or mutually exclusive processes, and one of them is Benford distributed, the resulting effective stochastic variable for the combined process is Benford distributed. This was referred to by Hamming [5,6] as the "persistence" property of the reciprocal distribution.

It is this combination of properties that prompted us to focus on the Benford distribution introduced above and in the phenomenological fits to data performed below.

Property (a) above follows from the general relationship existing between rank and size and the application of the inverse function theorem. Its proof can be found in [10].

Since $d(uv) = udv + vdu$, the error resulting from multiplying rounded numbers is dominated by the first significant digits of the numbers being multiplied. However, [11], as the Benford distribution favors the smaller first digits, it follows that the Benford distribution tends to produce less error than if the first digits were uniformly distributed. Property (b) then follows.

Property (c) is quite remarkable and is the consequence of a theorem in probability theory proven by T. Hill in 1995 [7,8], implying that the reciprocal or Benford distribution is in a sense "the distribution of distributions." Unlike the case of property (d) below, the proof of this deep theorem is technically quite demanding. It holds under very general conditions and explains a large body of facts known to apply to Benford distributed data. The theorem can be regarded as the analogue for pdf's of the situation in the Central Limit Theorem (CLT), which relates the law of large numbers for independent and identically distributed (iid) stochastic variables to the Gaussian distribution. In the CLT the superposition "of a large number of iid random variables with finite means and variances, normalized to have zero mean and variance 1, is approximately normally distributed." Here one has a large number of pdf's instead of iid stochastic variables. This theorem is general and applies to any stochastic system.

Property (d) is also remarkable. It follows from the fact that the arithmetic combination of two stochastic variables via multiplication, division, addition or subtraction involves the product or the sum of their respective pdf's. Carrying this out in detail [5,6] shows that if one of the pdf's is the reciprocal distribution, due to the properties of the integral of $1/x$, the resulting distribution for the combination is again the reciprocal.

In summary, the Benford distribution is then associated with geometrically distributed quantities, generates lower error rates in the combination of quantities than the uniform distribution, is "persistent" (or "contagious") and is the distribution for a random mixture of stochastic variables chosen at random.

## Results

### 1. Phenomenological Fits

A fit to the 953 data in the domain of Bacteria reveals that the functional form

$$y_{ORF}^{\text{Bacteria}}(G) = y_{ORF}^{B,0} + A_B \cdot G , \qquad (3)$$

with $y_{ORF}^{B,0} = 27 \pm 9$ ORF and $A_B = 0.943 \pm .005$ ORF/kbp provides an excellent fit with an $r$-parameter of 0.988. (Unless otherwise explicitly stated, $G$ will be expressed in kilobasepairs. Refer to the Methods Section of the paper and the Material S1 for a description of fitting methods and assumptions.)

A similar fit to the 69 Archaea data yields

$$y_{ORF}^{\text{Archaea}}(G) = y_{ORF}^{A,0} + A_A \cdot G, \qquad (4)$$

with $y_{ORF}^{A,0} = 231 \pm 57$ ORF and $A_A = 0.946 \pm .030$ ORF/kbp and an $r$-parameter of 0.967.

The 106 data for the Eukaryotes are well represented when fit to

$$y_{ORF}^{\text{Eukaryotes}}(G) = y_{ORF}^{E,0} + A_E \cdot \ln(1 + G/B_E) , \qquad (5)$$

with $y_{ORF}^{E,0} = -95 \pm 140$ ORF, $A_E = 3926 \pm 310$ ORF, $B_E = 3717 \pm 860$ kbp and an $r$-parameter of 0.947. (A fit to the functional form $y_{ORF}^{\text{Eukaryotes}}(G) = y_{ORF}^{E,0} + A_E \cdot \ln(G/B_E)$ does not do as well, especially for the smaller Eukaryotes.)

It is worth noting several features of these data and fits as plotted in Figure 1. We see

(i) that the Archaea (A) and Bacteria (B) line up together on a common curve, and that most of the Archaea are in the central portion of a joint (with the B's) Prokaryote (P) line. However, since the A and B fits share the same functional form, Eqns. (3) and (4) with equivalent slopes, we can also fit their combined data. This yields

$$y_{ORF}^{\text{Prokaryotes}}(G) = y_{ORF}^{P,0} + A_P \cdot G , \qquad (6)$$

where $y_{ORF}^{P,0} = 35 \pm 9$ ORF and $A_P = 0.946 \pm .005$ ORF/kbp are the values corresponding to a best fit with $r = 0.987$. (The fact that Archaea lie in the **central** portion of the Prokaryote (P) line could be partly due to bias in organism selection when the sequencing was done. We attribute significance to the fact that Archaea and Bacteria line up along the diagonal of the plot.)

One also notices that

(ii) the quantity $y_{ORF}^{(\min)}$ for Prokaryotes corresponds to $y_{ORF}^{\text{Prokaryotes}}(G^{(\min)})$ in Eqn. (6). But the tiny value of the straight-line offset, $y_{ORF}^{P,0}$, strongly suggests that the slope term, $A_P \cdot G^{(\min)}$, dominates and therefore without an additional scale to discriminate between the two terms in that equation it is impossible to produce an estimate for $y_{ORF}^{(\min)}$. However, as will be seen in Results Section C, a combined analysis of all data provides such a scale and leads to a bound for Eukaryotes, $y_{ORF}^{(\min)} < 167 \pm 400$ ORF. This is dominated by the slope term with the large uncertainty being dominated by the uncertainties for Eukaryotes. If we assume that the $G^{(\min)}$ values for the two domains are comparable, we can expect a bound for Prokaryotes commensurate with that for Eukaryotes given just above. In Eqn. (24) below we estimate $G^{(\min)}$ to be $90 \pm 130$ kbp, which is consistent with the arguments above.

(iii) At the other end of the **essentially** straight line where Prokaryotes lie, their genome size is limited to roughly 10–13 megabasepairs (Mbp). As will be seen below, this feature can be approximately understood in "simple" terms and may be related to the fact that Prokaryotes are not equipped (complex enough) to deal with the issues of non-linearity in the coding that would be needed in order to maintain a longer (and, therefore, in principle a more "capable") genome.

Finally, we notice that

(iv) while the smaller-genome Eukaryotes (E) are very close to the P-line (including some genome sizes smaller than the largest Prokaryotes), as $G$ grows the E's **begin to depart** noticeably from the straight line and linear regime into a non-linear regime well characterized for all Eukaryotes by $y_{ORF} \propto \ln(G/B)$. This occurs at around the same value where the ratio $ncDNA/cDNA$ becomes $\geq 1$ (cf. [2,12,13]). Furthermore, although we do not show it in our plot, the data [3,2,12,13] clearly indicate that for those Eukaryotic genomes where the $ncDNA$ fraction is quoted in the databases, one sees the well-known fact that most of the genomic material is in the form of $ncDNA$, which (eventually) dominates by orders of magnitude over the *cDNA* component as the total genome size increases.

The features listed above can be accommodated, accounted for and unified in a natural way by the properties of the Benford (or reciprocal) distribution if the pdf for ORFs as a function of full genome size is the Benford distribution. We briefly argue and motivate this in the next subsection.

## 2. The Origin of the Benford Distribution for the Distribution of ORFs in a Genome as a Function of Full Genome Size

**2.1. The cell environment is stochastic.** The detailed processes occurring within a cell are stochastic, as they take place in a complex and noisy physico-chemical environment [14–16]. Then, just as for any ensemble of molecules (or any other many–body system) genome dynamics can, in principle, be described by statistical mechanics [16], where the central object for the description of the dynamics is the probability distribution function (pdf) for the stochastic variable involved in the process under study [17].

For the genome, this pdf must take into account the many (and very different) processes that occur under the control of the genome, including how the nature of those processes changes as a function of genome size or as time elapses and/or both the internal and external environments change. We also note that in general, there is a positive correlation between genome size and organismic complexity or functions, so that a larger genome size brings with it the potential for more functions. This will affect how the larger number of base pairs contribute to the effective dynamics of ORFs and, therefore, their pdf as a function of full genome size $p_{ORF}(c+nc)$.

**2.2 ORFs (Open Reading Frames).** ORFs are **defined** [18,19] as stretches of genome DNA base pairs contained between a start and a stop codon. They contain **all** of the *cDNA* in the genome, but may also contain some pieces of *ncDNA*, such as introns, which (primarily) in Eukaryotes intersperse the stretches of *cDNA* between the ORF's start and stop codons.

An ORF is the part of a protein-coding gene that is translated into protein. ORFs are present in genomes from all three domains of Life. ORFs have special physical relevance due to the fact that they carry all of the *cDNA* in the genome. ORFs are first converted to pre-*mRNA* and then further converted to *mRNA* in the process of transcription. Eventually the information originally present in ORFs is translated into functional *RNA* and proteins. (We note that since ORFs contain all the *cDNA* in the genome they carry **at least** as much information as is contained in the *cDNA* fraction of the genome).

Of course the transmission of information from ORFs to *mRNA* and on to protein and functional *RNA* is an extremely complicated process. However, for our present purposes where we are interested in their overall effective pdf, it is only necessary to notice (independent of the specific details) that the processes of evolution in the nature of biological function must be closely related to the evolution of the number of ORFs within a genome.

**2.3. ORF phenomenology differs between Prokaryotes and Eukaryotes. The number of ORFs is also sensitive to the amount of *ncDNA* in the genome.** Although ORFs are present in all domains of Life, their phenomenology in Prokaryotes and Eukaryotes is markedly different. For example, in relation to genome size (and its various components), it is known that for Prokaryotes the number of ORFs per mRNA is **greater** than 1, whereas for Eukaryotes the number of ORFs per mRNA is, on average, only **slightly larger** than 1. In Prokaryotes several ORFs with their proteins are transcribed into a single mRNA. In Eukaryotes one ORF corresponds essentially to one mRNA [18].

It is also well known that the ratio of the *cDNA*–component to the *ncDNA*–component in a genome differs substantially between Prokaryotes and Eukaryotes. One observes (see, for example, [1,12,13]) that

$$\left(\frac{cDNA}{ncDNA}\right)_{\text{Prokaryotes}} \gg \left(\frac{cDNA}{ncDNA}\right)_{\text{Eukaryotes}} \qquad (7)$$

for all but the largest Prokaryotes. Put together with the remark from the preceding paragraph, this implies that the probability distribution function for ORFs in a genome **must** depend not only on the amount of *cDNA* but also on the *ncDNA* fraction of the genome and how they are apportioned.

Therefore, in developing the pdf $p_{ORF}(c,nc)$, from which the number of ORFs in the genome can be computed, one must consider the coupled influence of *cDNA* and *ncDNA* in describing the functionality of ORFs. (The pdf for ORFs as a function of **only** the *cDNA* fraction of the genome was inferred for a species from each of the three Domains of Life in Ref. [20]. There it was found to be expressible in each case as the superposition of two mutually exclusive pdf functions.)

That is, ORF dynamics depends on **both** the *cDNA* and the *ncDNA* components of the genome and not only on the *cDNA* component.

**2.4. Evolution and the proliferation of gene functions.** During the course of evolutionary history there has been a proliferation of gene functions (as compiled, for example, in Gene Families). This is the result of the four forces of evolution [1] acting on living systems. Through both adaptive and non-adaptive processes evolution gives rise within the physico-chemical environment of the living system, its ecological neighbors (of the same or different species) and the external environment, to new genes that enable the necessary functions to support Life. That is, new ORFs are also generated.

The incorporation of new genes and gene functions into the genome is generally the result of the evolution of pre-existing genes and functions. The extra underlying base pairs come, for example, from the relatively frequent **duplication** of regions in the genome that may have contained one or more genes [21]. Thus, as evolution takes place there are additional base pairs, both as *cDNAs* and *ncDNAs* of different types, which bring with them the potential for new and/or enhanced (biological) functions in larger genomes (for example related to development and genomic regulation). A qualitative description of some of the most important known properties of *ncDNA* is available in many current textbooks such as References [19,22].

But as genome size increases (in addition to the above) one can anticipate that the increase in genome length will eventually induce non–linearities as well as generate an enrichment of the interactions between the various DNA components and their functions in some non–trivial ways as more (biological) functions are implemented (see refs. [23], [24] and [12,13]). Furthermore, biological functions are usually related to domains in proteins (segments of ORFs), which during genome evolution are subject to rearrangement. In fact, domains act as modules and, in Eukaryotes, they are particularly combined into multiple forms in multi–domain proteins [25]. This rearrangement, particularly enhanced in Eukaryotes, could synergistically increase the functional repertoire encoded in the genome and therefore involve non–linear information control.

In view of this, it is therefore reasonable to infer that a larger genome size entails a concomitant **increase** in

(a) the number of biological functions that the living system can potentially express, as well as in

(b) the number of chemical processes that are controlled by the genome.

Since ORFs and genes are closely related, fulfilling these requirements has important consequences for the statistical mechanics of ORFs and hence their pdf. In this context, (a) implies an increase in the number of distinct pdfs for the base pairs that must combine in order to account for the statistical mechanics of the genome, while (b) implies an increase in the probability of errors taking place as a result of non-linearity, error combination and accumulation. The latter expresses the biological fact that segments of ORFs (domains) combine to produce further complexity, without recourse to *de novo* ORF invention.

**2.5. Putting the above together.** We have seen that during evolution the combination of adaptive and non-adaptive processes leads to changes in genome size that accompany deep modifications to living systems. These functional changes affect gene number, total genome length and the proportion of cDNA-to-ncDNA packed inside the genome, which may be expected to lead to speciation and beyond.

All these changes of course naturally have a reflection on the statistical mechanics of the genome. And more specifically on $p_{ORF}(c,nc)$, the pdf for ORF number as a function of full genome size. This function reflects the changes taking place in the stochastic dynamics of the genome albeit at a very general level. However, in view of properties (a) and (c) listed in Introduction Section B, the combination of gene duplication and the emergence of new functions during the evolutionary process can be incorporated into ORF phenomenology if $p_{ORF}(c,nc)$ is related to the Benford distribution.

Furthermore, once the Benford distribution for ORFs sets in, its persistence (property (d)) ensures that it will dominate other effective distributions (e.g., uniform, normal or Poisson) while, in view of property (b), also mitigating the size of errors in the combination of functions.

Thus multiple lines of argument lead to the same conclusion: the distribution of ORFs as a function of genome size, $G$, should follow a Benford distribution.

**2.6. A proposal for the pdf of the number of ORFs in a genome.** The above then suggests that we write

$$\tilde{p}_{ORF}(c,nc) = \frac{A}{G} = \frac{A}{c+nc}, \quad (8)$$

with $A>0$.

However, as written, the above pdf has two problems for any application to living systems: (1) it diverges as $G \to 0$ and (2) it does not accommodate the fact that living systems **cannot exist** for arbitrarily small genome size. These two difficulties can be addressed by a straightforward generalization of the form given in Eqn. (8) for the Benford pdf, namely, by introducing a new $p_{ORF}(c,nc)$ defined as

$$p_{ORF}(c,nc) = \frac{A}{B+(c+nc)}, \quad (9)$$

with $A, B>0$. The value of $y_{ORF}(G)$ is then given by

$$y_{ORF}(G) = y_{ORF}(G^{(min)}) + \int_{G^{(min)}}^{G} dg \cdot \frac{A}{B+g}, \quad (10)$$

where $y_{ORF}(G^{(min)}) \equiv y_{ORF}^{(min)}$ is the number of ORFs corresponding to the minimal genome size, $G^{(min)}$. Performing the integral in Eqn. (10) gives

$$y_{ORF}(G) = \left\{ y_{ORF}^{(min)} - A \cdot \ln(1+G^{(min)}/B) \right\} + A \cdot \ln(1+G/B) \quad (11)$$

for the number of ORFs in a genome of size $G \geq G^{(min)}$, and is identical to what we derived in Eqn. (2).

Dropping the statistically insignificant combination of the first two terms in Eqn. (11) gives the best fit to the data for Eukaryotes: $A = 4016 \pm 280$ ORF and $B = 4106 \pm 680$ kbp, which corresponds to a slope for small $G$ of $A/B = 0.978 \pm .100$ ORF/kbp, which is consistent with the slopes in Eqns. (3), (4) and (6).

Except for the region of the largest Prokaryotes, the dashed line in Fig. (1) shows that Eqn. (11) describes rather well the available data for the three domains of Life.

Next we will show how Eqn. (11) also interpolates between the genomic phenomenology of Prokaryotes and Eukaryotes.

## 3. Estimating the Size of the Region for the Split between Prokaryotes and Eukaryotes

A cursory look at Eqn. (11) shows that it describes two regimes: the first corresponding to $G^{(min)} < G < B$ and a second where $G^{(min)} < B < G$. In the (first) regime with $B>G$, expanding the logarithm to lowest order produces (here $\Delta y_{ORF}(G^{(min)}) \equiv y_{ORF}^{(min)} - A \cdot \ln(1+G^{(min)}/B)$).

$$y_{ORF}(G) \approx \Delta y_{ORF}(G^{(min)}) + \frac{A}{B} \cdot G + O[A(G/B)^2], \quad (12)$$

while in the (second) regime with $G>B$ we find

$$y_{ORF}(G) \approx \Delta y_{ORF}(G^{(min)}) + A \cdot \ln(G/B) + O[AB/G]. \quad (13)$$

The functional forms of these two regimes correspond to the two functional forms found earlier from fitting the data in Fig. (1), and quoted in Eqns. (5) and (6). We therefore infer that Prokaryotes are well described by the linear regime of the same form of pdf as the one that, in its non–linear regime, corresponds to the Eukaryotes. Furthermore, as mentioned earlier, the slope for the Eukaryotes in the linear regime is consistent with the slopes in Eqns. (3), (4) and (6) for the Prokaryotes.

The regime change takes place at genome sizes near $B$ and we can estimate the approximate position and size of this region. In fact, although we could equate Eqns. (11) and (6) and solve analytically for their point(s) of intersection, it is more useful to look for the **region** where a linear approximation to Eqn. (11) matches Eqn. (6).

The region of genome sizes beyond which $y_{ORF}(G)$ departs from a straight line and the Prokaryotes and moves into the region where Eukaryotes lie defines a "branching region" between Prokaryotes and Eukaryotes.

A "branching point" with genome size $g_0$ that **formally** characterizes this region can be introduced into Eqn. (11) by adding and subtracting it in the argument of the logarithm,

$$A \ln\left(1 + \frac{G-g_0}{B} + \frac{g_0}{B}\right) =$$

$$A \ln\left(1+\frac{g_0}{B}\right) + A \ln\left(1+\frac{G-g_0}{B+g_0}\right). \quad (14)$$

From the last term in Eqn. (14) we see that there are two obvious (and mutually exclusive) $G$ regimes: $[P]$ (roughly corresponding to Prokaryotes) and $[E]$ (roughly corresponding to Eukaryotes)

$$[P]: \left|\frac{G-g_0}{B+g_0}\right|<1 \qquad [E]: \left|\frac{G-g_0}{B+g_0}\right|>1 \quad (15)$$

with a boundary in the region of $G_{P/E} = B + 2g_0$. This value also sets the scale for the maximum genome size of Prokaryotes to a few times B, or around 8–12 Mbp.

Expanding Eqn. (14) in regime $[P]$ around the (for now arbitrary) point $g_0$ and requiring that the calculated $y_{ORF}(G)$ match the fit line for Prokaryotes in Eqn. (6), we find that the various constants must be related as follows (and to simplify the notation and avoid clutter, in the following we have made the substitutions from Eqn. (6) that $y_{ORF}^{P,0} = a$ and $A_P = b$):

$$g_0 = \frac{A}{b} - B, \quad (16)$$

and

$$y_{ORF}^{(min)} = a + b \cdot g_0 + A \cdot \ln\left(\frac{1+G^{(min)}/B}{1+g_0/B}\right). \quad (17)$$

Since we took $G^{(min)}$ to be the minimal genome size, it follows that $G^{(min)} < g_0$. Then from Eqn. (17) we have that $y_{ORF}^{(min)} < a + b \cdot g_0$. A lower bound is clearly provided by $G^{(min)} = 0$ in the form $y_{ORF}^{(min)} > a + b \cdot g_0 - A \cdot \ln(1+g_0/B)$. Applying the results from the fits leads to $g_0 = 140 \pm 430$ kbp, $32 \pm 17 < y_{ORF}^{(min)} < 167 \pm 400$ ORF, and $G_{P/E} = 4385 \pm 290$ kbp. The large errors on many quantities are a direct reflection of the large scatter of data about the fit line for the Eukaryotes. The uncertainties associated with the results for Prokaryotes are much smaller. Nonetheless, we consider this result highly suggestive, especially in view of other estimates of the minimal genome size [26–30].

We also see that Eqn. (11) accommodates the main observed features listed above under (i), (ii), (iii) and (iv).

### 4. Bounding the Minimal Eukaryotic Genome

A bound on the minimum Eukaryotic genome size, $G_E^{(min)}$, can be produced by equating the phenomenological form in Eqn. (5) to the more general expression in Eqn. (11). This produces

$$y_{ORF}^{(min)} = y_{ORF}^{E,0} + A_E \ln\left(1+\frac{G_E^{(min)}}{B_E}\right) > 0, \quad (18)$$

where we have also implemented a minimal constraint on $y_{ORF}^{(min)}$. Solving this simple inequality for $G_E^{(min)}$ produces

$$G_E^{(min)} > B_E\left[\exp\left(-\frac{y_{ORF}^{E,0}}{A_E}\right) - 1\right] \approx -\frac{y_{ORF}^{E,0} B_E}{A_E}, \quad (19)$$

where the second form applies only if $y_{ORF}^{E,0} \ll A_E$ (which is the case). Using the fitted parameter values below Eqn. (3) produces the result

$$G_E^{(min)} > 90 \pm 130 \, kbp, \quad (20)$$

which is consistent with the estimate for g0 ($G_E^{(min)} < g_0$) produced in the previous section.

## Discussion

We now discuss the physical interpretation and some of the consequences of the above results. The key points on which this discussion is based are that there exist clearly differentiated patterns for ORF number vs. genome size in Prokaryotic and Eukaryotic genomes, **and** in the observation that, except for statistical fluctuations, all the genomes are below the red diagonal line in the figure.

Our interpretations rely on entropy arguments, and in particular the consequences of maximal entropy. The Gibbs entropy for classical physical systems and the Shannon entropy [31,32] used in Information Theory [33] share a common form (although with different constants, $k$): $H \equiv -k \cdot \sum_{i=1}^{W} p_i \log p_i$, where the $W$ states $i$ of a stochastic variable have (normalizable) probability distribution functions, $p_i$. Maximizing $H$ with respect to the $p_i$ requires that those $p_i$ be equally probable, or $p_i = 1/W$ for the normalized pdf. Thus $H^{max} = k \cdot \log(W)$, which only requires counting the states $W$ to deduce $H^{max}$. Of course, this is also the form of the Boltzmann entropy for closed physical systems in equilibrium, written in its usual notation: $S_{Boltzmann} = k_B \cdot \log W$, where $W$ is the number of equiprobable states and $k_B$ is Boltzmann's constant.

Because our focus is on genome information storage and transmission, Shannon entropy is most relevant. That entropy is interpreted as "the measure of information received when the actual value of the stochastic variable is observed" [34]. Alternatively, it can be interpreted (cf. [31,32], Section 7, pp. 53–54) as the rate at which information in the stochastic variable can be transmitted by the communication system. The maximum entropy of a variable then characterizes the **achievable maximum information** that can be carried by the variable.

In what follows we ignore the constant, $k$, which plays no essential role, and assume that we are treating equally probable events and maximal attainable entropy (simply denoted by $H$). A good example is provided by tossing a single fair die, which has $W = 6$ equally probable outcomes and thus a maximal entropy of $H = \log(6)$. If one tossed $N$ such dice and considered the tosses to be statistically independent, the number of states for the $N$ dice would then be $W = 6^N$ and maximal entropy $H = N \cdot \log(6)$.

For a genome of length $G$ one has that if each of the $G$ positions in the genome can be occupied by one of four nucleotides the number of (equiprobable) states is $W \propto 4^G$. (This is the exact analogue of the dice example given above. Of course, as is well known, this is only an idealization and an approximation. See, e.g. Ref. [35].) Therefore for genomes the **maximum entropy** is achieved when $H^{Genome} = (\log 4) \cdot G$. In the log-log plot of Fig. 3.1 this corresponds (using the right-hand axis label) to an appropriately shifted diagonal, parallel to the red line drawn there. This can be suitably interpreted as the maximum total information that a genome of length $G$ can handle. In other words, full genome size dictates its maximum achievable entropy. We see at once from the figure that to a good approximation, $y_{ORF}$ for Prokaryotes falls exactly on this line, and that Eukaryotes **clearly** have departed from this line for genome sizes beyond $\sim 4$ Mbp (around the size of the quantity B in the fits).

Information is, by definition, an additive quantity. For a living system (L/S) it resides in many places. Fundamentally, it is in the selection of components used by Life for their basic architecture and the way the L/S operates (which defines a Shannon entropy $H^{(0)}$, in the genes and ORFs ($H^{\text{ORF}}$) and in the ncDNA piece ($H^{\text{ncDNA}}$). (Naturally, each of the above locations for information can be split into many others for which the rules of information theory, in turn, apply.)

However, for the remainder of this discussion we will restrict ourselves to some of the most generic features of the information associated with the ORF-component of the L/S, ($H_{\text{L/S}}^{\text{ORF}}$), as they relate to Information Theory.

We examine the maximum value that $H_{\text{L/S}}^{\text{ORF}}$ can reach. This will happen when the pdf for the frequency of expression of the ORFs in a genome, $p_{\text{L/S}}^{\text{ORF}}(k)$, which describes the $k$ states that empower the information carried by the ORFs in the L/S, is the uniform distribution. That is, when $p_{\text{L/S}}^{\text{ORF}}(k) = \rho = \text{constant}$ for all $k$. That is, for maximum potentially achievable entropy we assume that each of the $y_{ORF}$ in a genome of size $G$, given in Eqn. (11), is equiprobable and independent of the others. Then, for each genome, the normalization of $p_{\text{L/S}}^{\text{ORF}}(k)$ trivially yields that $\rho = 1/y_{ORF}(G)$. Therefore, **under these assumptions**, the maximum Shannon entropy for the ORFs is given by $H_{\text{L/S}}^{\text{ORF}} = \log(y_{ORF}(G))$ and measures the **maximum information** that can be carried by the ORFs in a L/S whose genome size is $G$.

But Eqn. (11) has the two regimes identified in Eqn. (15) depending on whether $G < G_{P/E}$ or $G > G_{P/E}$. Thus, there also exist two different regimes for the information carried by ORFs in extant Life: one for Prokaryotes and another for Eukaryotes. They are roughly separated by the value of $G_{P/E}$, which as was found earlier has a magnitude consistent with what other authors (for example [36]) have identified using completely different arguments.

Using Eqn. (12) we find that for genomes where $G < B$,

1. and also such that $A \cdot G/(B \cdot \Delta y_{ORF}) < 1$ the maximum information in ORFs is $\text{Max} H_{\text{L/S}}^{\text{ORF}} = \log(y_{ORF}) \propto G$,
2. while if $A \cdot G/(B \cdot \Delta y_{ORF}) > 1$ then $\text{Max} H_{\text{L/S}}^{\text{ORF}} = \log(y_{ORF}) \propto \log G$.

That is, when $G < B$, the information carried by the ORFs in the smaller Prokaryotes (or small Eukaryotes) approaches the maximum value accessible to a genome of size $G$.

On the other hand for genomes where $G > B$, which includes the largest Prokaryotes and the vast majority of Eukaryotes, one finds using Eqn. (12) that

1. if $G$ is also such that $A/(B \cdot \Delta y_{ORF}) \cdot \log(G/B) < 1$ then the maximum information in ORFs is $\text{Max} H_{\text{L/S}}^{\text{ORF}} = \log(y_{ORF}) \propto \log G$,
2. while if $A/(B \cdot \Delta y_{ORF}) \cdot \log G > 1$ then $\text{Max} H_{\text{L/S}}^{\text{ORF}} = \log(y_{ORF}) \propto \log \log G$.

Therefore, since $G > \log G > \log(\log G)$ for all $G$, we see that the ORFs at maximum Shannon entropy are **always below** the maximum entropy that the full genome would potentially be capable of achieving.

The above has an interesting consequence. Since the information carried by the cDNA fraction of the genome is in the ORFs, the necessary complement for the genome to achieve its maximum information content for genomes with $G \geq B$ is clearly provided by the non-coding fraction of the genome that, when added to the coding piece of the genome represented by cDNA, saturates the information to $G$ from "only" $\log \log G$. In other words, the ncDNA component contributes to the entropy what the ORFs and, therefore, the cDNA fraction cannot. As a consequence, we see why the fraction of ncDNA has the potential to be much larger than the cDNA fraction, all of which is contained in the ORFs.

For genomes smaller than 8 or so Mbp in the linear regime for $y_{ORF}$, the cDNA fraction suffices to saturate the maximum information possible (actually to about 90%, as the fit to the Prokaryotes shows), but for longer genome sizes the contribution of spliceosomal introns and other forms of ncDNA is necessary in part to help support the function of ORFs and mostly in order to saturate the maximum information value possible for the genome.

This "information crisis" occurs for genomes with $G \sim B$, and provides a non-adaptive reason for the increase in genome size. It impacts on the fact that new opportunities arise for novel activities controlled by the ncDNA fraction but, since the ORFs are Benford distributed and property (b) of Introduction Section B applies, it also opens a new avenue for reducing the effective mutation rates in larger genomes. One can imagine that since the inverse of the product of the effective population $N_e$ with the mutation rate $u$ controls the power of genetic drift, a further reduction in $u$ originating in improved control due to Benford allows for a somewhat larger $N_e$ than the ones usually contemplated in the literature [36]. Phylogenetically and qualitatively speaking we see that a reduced mutation rate $u$ with genome size (due to Benford) is consistent with the observation that there is "an inverse relationship between organism size (and therefore genome length) and $N_e u$" [36].

## 1. Conclusions

In conclusion, the genomic data from all three domains of Life support the proposition that the ORFs belonging to a genome of $c + nc$ base pairs are distributed according to a "reciprocal" (or Benford) pdf. This observation helps unify and explain some salient features of the observed phenomenology, but needs some qualification. Eukaryotes are represented by the full non-linear regime of a Benford distribution for the number of ORFs as a function of $G$, while Prokaryotes correspond to the linear regime of the **same** (viz., they have consistent slopes) Benford distribution.

More specifically, Eqn. (11) accommodates the facts that in a plot of the number of ORFs in a genome vs. full genome size expressed in kbp, (i) the Archaea and Bacteria line up on a common curve fitted by Eqn. (4) (and consistent with Eqn. (11) in its linear regime), (ii) the **minimal** size of Prokaryotes is bounded from above by $167 \pm 400$ ORF, (iii) that there is a **maximal** Prokaryote size on the order 8–12 Mbp, below which the non-linear effects associated with the Benford regime are not felt, but beyond which (iv) the non-linearity of Eqns. (5) or (11) dominates and encompasses Eukaryotes with genome sizes approximately larger than 4–5 Mbp. In addition this distribution also allows one to compute a minimum genome size for Eukaryotes (which with the data used in this paper also turns out to have a large error) of about $90 \pm 130$ kbp.

In other words our results apply to all of Life, with Prokaryotes being in the linear regime of a Benford distribution, while Eukaryotes are in the non–linear regime of the same Benford distribution.

The Benford distribution depends only on the full genome size, $G$, and necessarily mixes cDNA and ncDNA fractions. We infer from this that the relative sizes of these two fractions must also depend on the full genome size, and therefore are not independent of each other: for a given genome of size $G$ the ratio of these two fractions is a function of the **full** genome size. If the average size of a Prokaryote ORF is 1 kbp (see below), then the values quoted for the linear slope parameters (viz., $A_B$ and $A_A$ in Eqns. (3) and (4)) indicate that the ncDNA fraction in most Prokaryotes is less than (or the order of) 10% of the total genome length.

By appealing to the most basic notions of information theory, we have seen that information is bounded in a genome of size $G$ and must lie below a maximum possible value of $\log(G)$.

If the average size of an ORF is of the order of 1 kbp in Prokaryotes (as is usually assumed [16,21]) the cDNA fraction (expressed as ORFs) almost saturates the maximum information that can be packed into the genome. But in Eukaryotes the situation is very different: the information in cDNA (ORFs) is far less than the information potentially contained in ncDNA, which could (at least in principle) be somehow expressed at a level that eventually saturates the upper bound mentioned above.

Whatever their specifics we see that the mechanisms for ORF–information management in Eukaryotes must involve non–linear information control. In Prokaryotes, on the other hand, the expression of the cDNA is essentially linear and therefore potentially simpler, albeit limited.

Finally we remark that information theory can provide us with a promising starting point for understanding and interpreting the results of our "Life-wide" fits.

## Methods

We give here a short description of the data and the methods we have used for obtaining and interpreting our fits.

### 1. Fits and Statistical Analysis

The data that we analyze were obtained from the completed entries (more than 1000) in the GOLD database [3] in early 2010. These data specify the genome size (in kilobasepairs: kbp), the number of ORFs identified, and the metadata (specifying the organism) for each organism. There were a number of typographical errors in the database that we corrected using the accompanying publication information. A few additional entries for recently sequenced Eukaryotes that had not yet been added to the database were made from the literature.

The 953 separate data for Bacteria ranged in size from 188 ORF for *Candidatus Hodgkinia cicadicola* to 9771 ORF for *Sorangium cellulosum*, a range of nearly two orders of magnitude. The 106 separate data for Eukaryotes ranged in size from 464 ORF for *Guillardia theta* to 50000 ORF for *Oryza sativa*, which is two orders of magnitude, in contrast to nearly four orders of magnitude in genome sizes. The 69 Archaea range from 643 ORF for *Nanoarchaeum equitans* to 4853 ORF for *Methanosarcina acetivorans*, a span slightly less than one order of magnitude.

In the Results Section we fit the number of ORFs to functions of the total genome size. In order to perform these fits we need a calculational scheme that takes into account appropriate uncertainties for the data that we use. We assume no uncertainty in the number of base pairs in each genome, and use this as our independent variable. The spread in the data as a function of the number of ORFs (or genome size) provides an estimate of the uncertainties associated with our data. It is shown in Figs. (S1) and (S2) of the SM that the data for Bacteria have small spread for small genome size and large spread for large size, which removes from consideration a uniform error (used in ordinary least-squares fitting). A similar result holds for Eukaryotes, as shown in Figs. (S3) and (S4) of the SM. We therefore assume that the uncertainty in the number of ORFs for each genome size is proportional to the number of ORFs in each datum, $y^i_{ORF}$. That is, we assume a common fractional uncertainty per ORF (denoted by $\lambda$), which is consistent with the spread in the fractional residuals from our fits (see Figs. (S2) and (S4) of the SM, and the discussion in the next paragraph). Denoting the uncertainty in the $i$-th datum by $\sigma_i = \lambda y^i_{ORF}$, the fitting is done by minimizing the usual $X^2$ function with respect to all parameters in the fitting function, $F$:

$$\chi^2 = \sum_{i=1}^{N} \left( \frac{y^i_{ORF} - F_i}{\sigma_i} \right)^2, \quad (21)$$

where $N$ is the number of data and $F_i$ is the value of the fitting function corresponding to the $i$-th datum. This is a form of weighted least-squares minimization (an extensive discussion of this is provided in the SM). Although the parameter $\lambda$ plays no role in the determination of the best-fit parameters, it is required for estimating uncertainties in those parameters and we used the maximum likelihood estimate for $\lambda$ (as described in detail in the SM). Note that the maximum likelihood estimate for $\lambda$ is identical to adjusting the fitted value of $X^2$ per degree of freedom to 1 (cf. the SM).

Had we resorted to ordinary least-squares minimization (corresponding to uniform $\sigma_i$), the large range of ORF sizes would have rendered the fits sensitive only to the largest genomes, and therefore largely useless. In contrast an important consequence of our form of $\sigma_i$ is that the fits are equally sensitive to all data, large or small. We emphasize that for any fit the resulting distribution of fractional residuals $[(y^i_{ORF} - F^{fit}_i)/\sigma_i]$ for any genome size should be roughly independent of that size (e.g., a uniform variance). This is demonstrated in the in Figs. (S1) and (S2) for Bacteria, in Figs. (S3) and (S4) for Eukaryotes, and the extensive accompanying discussion. We also tested the fractional residuals against the hypothesis that they were Gaussian distributed. Outliers for the Bacteria case (defined as being more than three standard deviations from the fit) were too many (roughly 2% compared to an expected 1/4%) to achieve a good P-value. The effect of the outliers was tested by deleting them and refitting, resulting in fits that were statistically equivalent to the full fits that were reported in the Results Section. Other than the excessive number of outliers the Gaussian comparison was quite satisfactory; the SM should be consulted for details.

In the Results Section we quoted standard deviations for the fitted parameters, rather than P-values. The latter can be deduced from the t-statistic, which is the ratio of a parameter's value to its standard deviation. Any value greater than about 2 satisfies the usual 95% criterion for a significant result (viz., statistical fluctuations alone cannot account for the fit). The quality of our fits (for at least one fit parameter and thus for the fit as a whole) generates t-statistics and F-statistics that are far larger than needed to satisfy this criterion. The corresponding P-values are tiny and thus uninformative and have not been quoted in the text (details are in the SM).

## Supporting Information

**Figure S1 Bacteria Data, Fits and Residuals.** Distribution of Bacteria data about the fit line (in red) and absolute residuals relative to that line (in green) as a function of genome size in kbp. Fractional residuals are shown in blue.
(TIFF)

**Figure S2 Bacteria Fractional Residuals.** Distribution of Bacteria fractional residuals (in blue) as a function of genome size in kbp. The black line corresponds to the fit, while the red dashed lines are one standard deviation away, and the dotted black lines are two standard deviations away.
(TIFF)

**Figure S3  Binned Eukaryota Fractional Residuals.** Eukaryota fractional residuals (in black) sorted into 7 one-standard-deviation-wide bins compared to an assumed Gaussian distribution with the same mean and variance (in red).
(TIFF)

**Figure S4  Eukaryota Fractional Residuals.** Distribution of Eukaryota fractional residuals (in blue) as a function of genome size in kbp. The black line corresponds to the fit, while the red dashed lines are one standard deviation away. The dotted magenta line at 0.285 is the mean of the fractional residuals in units of (one) standard deviation, while the other two dotted magenta lines are one standard deviation away from the mean.
(TIFF)

**Material S1  Material on Fits, Fitting Procedures, and Statistics.** Detailed notes and derivations.
(PDF)

## Author Contributions

Conceived and designed the experiments: JP-M JF TG. Analyzed the data: JP-M JF TG. Contributed reagents/materials/analysis tools: JP-M JF TG. Wrote the paper: JP-M JF TG.